\documentclass[sigconf,screen]{acmart} 

\usepackage{amsmath,amsfonts}
\usepackage{algorithmic}
\usepackage{graphicx}
\usepackage{textcomp}
\usepackage{xcolor}

\usepackage{textcomp}
\usepackage{xcolor}
\usepackage{array}
\usepackage{xspace}
\usepackage{listings}
\usepackage{longtable}
\usepackage{lscape}
\usepackage{threeparttable}
\usepackage{multirow}

\usepackage[normalem]{ulem}
\usepackage{pifont}

\AtBeginDocument{%
  }

\setcopyright{acmlicensed}
\copyrightyear{2018}
\acmYear{2018}
\acmDOI{XXXXXXX.XXXXXXX}

\acmConference[EASE 2025]{The 29th International Conference on Evaluation and Assessment in Software Engineering}{17–20 June, 2025}{Istanbul, Turkey}

\acmISBN{978-1-4503-XXXX-X/18/06}

\DeclareMathOperator*{\argmax}{arg\,max}

\newcommand{\rqone}{\textbf{RQ$_1$}: \emph{
How effective is \tool in addressing cold-start recommendation compared to state-of-the-art approaches?}}

\newcommand{\rqtwo}{\textbf{RQ$_2$}: \emph{How does \tool perform compared to state-of-the-art methods in mitigating popularity bias?}} 

\newcommand{\rqthree}{\textbf{RQ$_3$}: \emph{What is the effectiveness of the proposed reinforcement learning algorithm in an interaction-split recommendation setting?}}

\newcommand*{\tool}{\texttt{RAILS}\@\xspace}
\newcommand*{\LGCN}{\texttt{LightGCN}\@\xspace}
\newcommand*{\CR}{\texttt{CrossRec}\@\xspace}
\newcommand*{\LR}{\texttt{LibRec}\@\xspace}

\newcommand*{\GR}{\texttt{GRec}\@\xspace}

\newcommand*{\RLGCN}{\texttt{RL--LightGCN}\@\xspace}
\newcommand*{\RLBC}{\texttt{RL--LightGCN--BCLoss}\@\xspace}

\begin{document}

\title{Defusing Popularity Bias in Third-Party Library Recommendations: A Reinforcement Learning Approach}

\title{Bake Two Cakes with One Oven: Defusing Popularity Bias and Cold-start in Third-Party Library Recommendations with Reinforcement Learning}

\title{Bake Two Cakes with One Oven: Defusing Popularity Bias and Cold-start in Third-Party Library Recommendations with RL}

\title{Bake Two Cakes with One Oven: RL for Defusing Popularity Bias and Cold-start in Third-Party Library Recommendations}


\author{Minh Hoang Vuong}
\email{minh.vh210590@sis.hust.edu.vn}
\affiliation{%
	\institution{Hanoi University of Science and Technology, Vietnam}
	\city{Hanoi}
	\country{Vietnam}
}

\author{Anh M. T. Bui}
\authornote{Anh M. T. Bui is the corresponding author}
\orcid{0000-0001-7877-9438}
\email{anhbtm@soict.hust.edu.vn}
\affiliation{%
	\institution{Hanoi University of Science and Technology, Vietnam}
	\city{Hanoi}
	\country{Vietnam}
}

\author{Phuong T. Nguyen}
\email{phuong.nguyen@univaq.it}
\orcid{0000-0002-3666-4162}
\small\affiliation{%
	\institution{Universit\`a degli studi dell'Aquila}
	\city{67100 L'Aquila}
	\country{Italy}
}

\author{Davide Di Ruscio}
\email{davide.diruscio@univaq.it}
\orcid{0000-0002-5077-6793}
\affiliation{%
	\institution{University of L'Aquila}	
	\city{67100 L'Aquila}	
	\country{Italy}	
}


\begin{abstract}
Third-party libraries (TPLs) have become an integral part of modern software development, enhancing developer productivity and accelerating time-to-market. However, identifying suitable candidates from a rapidly growing and continuously evolving collection of TPLs remains a challenging task. TPL recommender systems have been studied, offering a promising solution to address this issue. They typically rely on collaborative filtering (CF) which exploits a two-dimensional {\it project-library} matrix ({\it user-item} in general context of recommendation) when making recommendations. We have noticed that CF-based approaches often encounter two challenges: (i) a tendency to recommend popular items more frequently, making them even more dominant, a phenomenon known as {\it popularity bias}, and (ii) difficulty in generating recommendations for new users or items due to limited user-item interactions, commonly referred to as the {\it cold-start} problem. 

In this paper, we propose a reinforcement learning (RL)-based approach to address popularity bias and the cold-start problem in TPL recommendation. Our method comprises three key components. First, we utilize a graph convolution network (GCN)-based embedding model to learn user preferences and user-item interactions, allowing us to capture complex relationships within interaction subgraphs and effectively represent new user/item embeddings. Second, we introduce an aggregation operator to generate a representative embedding from user and item embeddings, which is then used to model cold-start users. Finally, we adopt a model-based RL framework for TPL recommendation, where popularity bias is mitigated through a carefully designed reward function and a rarity-based replay buffer partitioning strategy. 
We conducted experiments on benchmark datasets for TPL recommendation, demonstrating that our proposed approach outperforms state-of-the-art models in cold-start scenarios while effectively mitigating the impact of popularity bias.
\end{abstract}





\maketitle




\section{Introduction}
\label{sec:introduction}

Third-party libraries (TPLs) play a crucial role in modern software development, enabling efficient and effective code reuse~\cite{ouni2017search,saied2018improving}. Their integration enhances the productivity and contributes to overall software quality improvement~\cite{li2021embedding}. 
Thung et al. conducted an empirical study revealing that 93.3\% of modern open-source software projects incorporate TPLs, with an average of 28 libraries per project~\cite{thung2013automated}.
A large-scale study of Android applications on Google Play found that, on average, 60\% of a mobile application's source code originates from external libraries~\cite{ma2016libradar}.
The widespread popularity and benefits of TPLs have resulted in the creation and publication of a vast number of libraries. As a result, developers face considerable challenges in identifying the most suitable TPLs for their software projects. 
Indeed, a recent study has shown that the number of available libraries is growing exponentially~\cite{saied2018improving}. As a result, searching for relevant software libraries can be a tedious and time-consuming task, potentially affecting developers' productivity. Moreover, external libraries are chosen based on various factors such as functionality, security, and performance, making it essential to find the right combination of TPLs for a project. Consequently, developers urgently need effective and efficient support in identifying suitable TPLs.

Recent research has recognized a substantial body of work on TPL recommender systems aimed at assisting developers in finding the right libraries for their projects~\cite{thung2013automated,ma2016libradar,ouni2017search,nguyen2020crossrec,li2021embedding}.
Most existing library recommendation approaches are based on collaborative filtering, a widely used technique that predicts user-item interactions using historical data.  
LibRec~\cite{thung2013automated} is one of the earliest CF-based library recommendation systems, combining collaborative filtering with pattern rules to suggest TPLs for Java projects.  
CrossRec~\cite{nguyen2020crossrec} leverages project similarity and adjusts library weights based on term frequency and inverse document frequency, demonstrating a high success rate in recommendations. 
These approaches primarily rely on identifying similar projects and/or libraries to generate recommendations.
LibSeek~\cite{he2020diversified} adopts a different CF-based approach, focusing on training a factor model using existing project-library interactions. Specifically, it utilizes Matrix Factorization, mapping projects, libraries, and project-library usage data into a shared latent space, where each project and library is represented as a vector of latent features.

Although CF-based approaches have shown promising results and have been widely adopted across various recommendation domains, including e-commerce, news portals, and software engineering, they come with certain limitations. Research has indicated that the conventional training paradigm, which focuses on fitting a recommender model to recover user behavior data, can introduce bias toward popular items~\cite{he2020diversified,wei2021model,gao2023alleviating,liu2023popdcl}. As a result, the model tends to favor recommending popular items while overlooking less popular ones that users might find valuable. This bias not only impairs the model’s ability to accurately capture user preferences but also reduces the diversity of recommendations.
Moreover, while CF-based methods aim to capture user behavioral similarities from historical user-item interactions to effectively infer user preferences, they struggle with the cold-start problem. This challenge arises when interactions for new users or new items are extremely sparse or entirely absent from historical data, making accurate recommendations difficult~\cite{zhu2019addressing,lu2020meta,wei2021contrastive}.

In this paper, we propose a reinforcement learning (RL)-based approach to tackle both popularity bias and the cold-start problem in TPL recommendation, \tool--Graph-powered \textbf{R}einforcement learning fr\textbf{A}mework for pop\textbf{I}larity-aware and co\textbf{L}d-\textbf{S}tart recommendation. RL has been extensively studied and has shown promising results in recommendation systems~\cite{zou2019reinforcement,afsar2022reinforcement}. Ideally, a recommendation policy should aim to maximize user satisfaction over the long term. By making sequential decisions and taking actions toward a long-term objective, an RL agent provides a more effective strategy for capturing user-item interactions~\cite{tang2019reinforcement}. Our proposed approach consists of three components. 
First, rather than depending solely on a two-dimensional user (project)-item (library) matrix for project/library collaborative embeddings, we employ a graph convolution network (GCN)-based model. This approach enables learning project preferences and project-library interactions, capturing more intricate relationships within interaction subgraphs.
Second, we introduce an aggregation operator to generate a representative embedding from project/library embeddings, allowing to effectively representing newly introduced project and library embeddings, particularly in cold start scenarios with limited historical interactions.
Finally, we adopt an RL-based framework for TPL recommendation, where popularity bias is addressed through a reward function that incorporates a popularity-debiased term and a rarity-based replay buffer partitioning strategy.
We conducted experiments on a benchmark dataset for TPL recommendation to evaluate our proposed approach against two state-of-the-art models, \CR~\cite{nguyen2020crossrec} and \LR~\cite{thung2013automated}. The comparison focused on the ability to handle cold-start scenarios and mitigate popularity bias. Experimental results show that our method achieves promising performance and proves highly effective in generating recommendations.
To facilitate future research, we publish a replication package including the dataset and source code of our proposed approach.\footnote{https://huggingface.co/GRAPHICS-repo/GRAPHICS}

\section{Proposed Approach}
\label{sec:proposal}

\begin{figure*}[!t]
\includegraphics[scale=0.5]{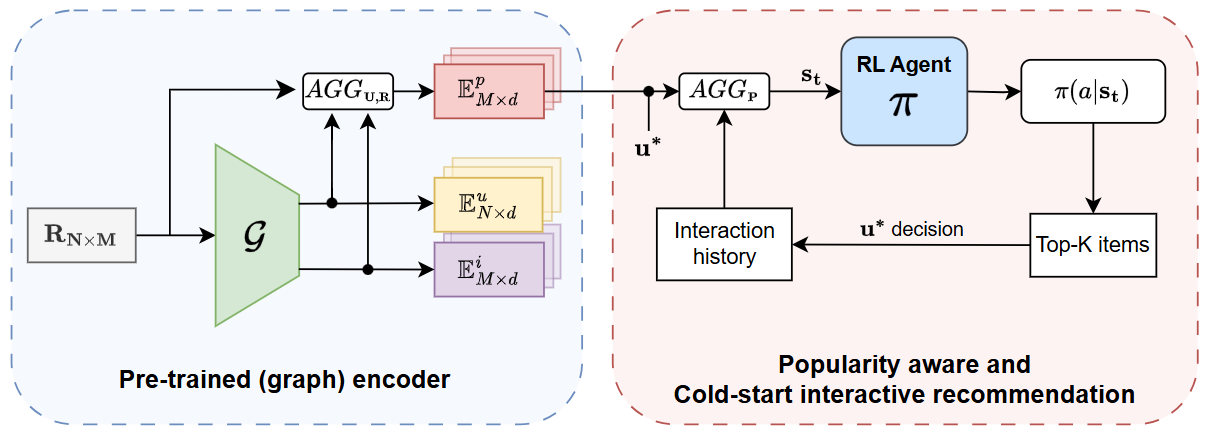}
\caption{The proposed framework}
\label{fig:framework}
\end{figure*}

This section describes our proposed approach to TPL recommendation, which is depicted in Figure~\ref{fig:framework}. We first formulate the problem and then describes three components of our proposed approach: (i) Collaborative Embedding (ii) Cold-Start representation and (iii) RL Agent for TPL recommendation. 

\subsection{Problem Formulation}
In the context of TPL recommendation, a project can be regarded as a user, while a library can be considered an item.
Let \(\mathcal{U} = \{u_1, u_2, \dots, u_N\}\) represent the set of \(N\) projects and \(\mathcal{I} = \{i_1, i_2, \dots, i_M\}\) represent the set of \(M\) libraries. The historical user-item interaction matrix is constructed by capturing the library usage of each project. We define \(\mathcal{R}^{N \times M}\) as the two-dimensional interaction matrix, where \(r_{kj} = 1\) if project \(u_k\) utilizes library \(i_j\), and \(r_{kj} = 0\) otherwise.

In order to learn the representation vectors of each project and library, we define an embedding model with three components: (i) a project encoder $\psi : \mathcal{U}\to\mathbb{R}^d$; (ii) a library encoder $\phi : \mathcal{I}\to\mathbb{R}^d$ where $d$ is the dimension of embedding vector and (iii) a scoring function $f : \mathbb{R}^d\times\mathbb{R}^d\to\mathbb{R}$. 
Given a project-library interaction pair $(u,i)$ ($u \in \mathcal{U}, i \in \mathcal{I}$), the likelihood of $i$ being chosen by $u$ is given by:
\begin{equation}
    y(u, i) = f(\psi(u), \phi(i))
\label{eq:cf-likelihood}
\end{equation} 
The embedding model will be trained with the training dataset and then will be employed to predict the projects' preferences towards libraries.

\subsection{Collaborative Embedding}
We refer to the representation vectors of projects and libraries as collaborative embeddings. A substantial body of research focuses on these embeddings, including ID embeddings in matrix factorization~\cite{zhou2011functional}, interaction graph embeddings in \texttt{LightGCN}~\cite{he2020lightgcn}.
In this paper, we adopt \texttt{LightGCN}~\cite{he2020lightgcn} as our collaborative embedding model, as it effectively leverages the complex graph structure of user-item interactions and encodes this information into the final embeddings. 
Let \(\mathcal{G} = (\mathcal{V}, \mathcal{E})\) represent the undirected project-library interaction graph, where the vertex set is defined as \(\mathcal{V} = \mathcal{U} \cup \mathcal{I}\). An edge \(\{u, i\}\) is included in the edge set \(\mathcal{E}\) if project \(u\) has interacted with library \(i\). 
We train the graph $\mathcal{G}$ with the contrastive based loss function, \texttt{BCLoss}, proposed by Zhang et al.~\cite{zhang2022incorporating}, which has been designed to alleviate the effect of popularity bias when training \texttt{LightGCN}. 

Let $\mathbb{E}^u_{N\times d}$ and $\mathbb{E}^i_{M\times d}$ be the trained project and library embeddings. We denote $e_u = \psi(u)$ and $e_i = \phi(i)$ as the embedding vectors of project $u$ and library $i$. Equation~\ref{eq:cf-likelihood} can be rewritten as $y(u,i)=f(e_u, e_i)$. We simplify this function by utilizing the cosine similarity. The likelihood of a pair $(u,i)$ will be computed as in Equation~\ref{eq:similarity}.
\begin{equation}
    y(u, i) = \frac{\overrightarrow{e_u} \cdot \overrightarrow{e_i}}{{\lVert\overrightarrow{e_u}\rVert\times\lVert\overrightarrow{e_i}\rVert}} 
\label{eq:similarity}
\end{equation}
Without loss of generality, let $\lVert\overrightarrow{e_u}\rVert = \lVert\overrightarrow{e_i}\rVert = 1\;\forall i,j$. Equation~\eqref{eq:similarity} can then be written as:
\begin{equation}
    y(u, i) = \overrightarrow{e_u} \cdot \overrightarrow{e_i}
\label{eq:normalized-similarity}
\end{equation}
This equation can be seen as the rating score of a pair of project-library $(u,i)$.

\subsection{Cold-Start Representation}
To effectively address the cold-start problem for new users (projects), we propose a representative embedding to encode a newly introduced project $u^*$ when it enters the system.

We first define the \textbf{project-based representative embedding} of library $i$, denoted by $\mathbf{p_i}$ as in Equation~\ref{eq:i-representative}.
\begin{equation}
    \mathbf{p_i} = \dfrac{\sum_{u\in\mathcal{U}^i}y(u, i)e_{u}}{\sum_{u\in\mathcal{U}^i}y(u, i)}
    \label{eq:i-representative}
\end{equation}
where $\mathcal{U}^i\subset\mathcal{U}$ is the set of projects that had interactions with library $i$ and $y(u, i)$ is the similarity measure of project $u$ and library $i$ as defined in Equation~\eqref{eq:similarity}. 

Supposing that $\mathcal{I}_{u^*}$ denotes a very short list of libraries having interactions with $u^*$, the cold-start representation of $u^*$ is determined using an aggregation operator:
\begin{equation}
    e_{u^*} = Agg(\mathcal{I}^{u^*}) = \dfrac{1}{|\mathcal{I}^{u^*}|}\sum_{i\in\mathcal{I}^{u^*}}\mathbf{p_i}
\label{eq:cold-start-repr}
\end{equation}
where $|\mathcal{I}^{u^*}|$ represents the size of $\mathcal{I}^{u^*}$ and $\mathbf{p_i}$ is the representative embedding of $i$.



It is important to note that many long-tailed items appear only once in the interaction matrix, meaning some items are linked to just a single user. In such cases, $\mathbf{p_i} = e_u$. As a result, multiple long-tailed items associated with the same user will have the same embedding as \(e_u\). This issue can be alleviated by incorporating some ``noise'' into \(\mathbf{p_i}\). To achieve this, we use the item embedding \(e_i\) as the noise component and introduce a weight \(\lambda\) to balance the contributions of user-based and item-based information in the final representation (Equation~\ref{eq:weighted-repr}).

\begin{equation}
    \mathbf{p_i} = \lambda\dfrac{\sum_{u\in\mathcal{U}^i}y(u, i)e_{u}}{\sum_{u\in\mathcal{U}^i}y(u, i)} + (1-\lambda)\mathbf{e_i}
\label{eq:weighted-repr}
\end{equation}

\subsection{RL Agent for TPL Recommendation}
Using RL to resolve a problem is often formulated as a Markov Decision Process (MDP), denote as $M = (\mathcal{S}, \mathcal{A}, \mathcal{T}, r, \gamma)$ where $\mathcal{S}$ and $\mathcal{A}$ represent the state space and the action space. $\mathcal{T}(s, a, s') = P(s_{t+1} = s'|s_t = s, a_t = a)$ is the transition probability from $(s,a)$ to $s'$. $r(s,a)$ represents the reward of taking action $a$ at the state $s$ and $\gamma$ is the discount factor. 
We formulate a RL-based recommender system as an offline RL task~\cite{kumar2020conservative} where the transition probability $\mathcal{T}$ and reward function $r$ are predicted by an offline model which has been trained on an offline dataset $\mathcal{D}$. In the context of TPL recommendation, we train the RL policy on the pre-trained embeddings $\mathbb{E}^u$ and $\mathbb{E}^i$.

At each recommendation step \( t \), the RL agent is in the current state \( s = u_t \). The system then recommends a library as an action \( a_t \) based on a policy \( \pi \). Following this recommendation, a reward \( r_{t+1} \) is computed. The policy \( \pi \) is trained to maximize the cumulative reward over \( \Delta T \) time steps, which represents the maximum number of recommendation steps. We utilize a \texttt{DQN}~\cite{van2016deep} as the backbone for training the recommendation policy with $\widehat{Q}(\mathbf{s}, a, \theta)$ be the state-action value function parameterized by $\theta$. 
The objective is to approximate the optimal state-action value function $Q^*(\mathbf{s}, a)$, which represents the expected cumulative reward an agent can obtain when starting from state $s$, taking action $a$, and subsequently following the optimal policy $\pi^*$ (Equation~\ref{eq:greedy-policy}).
\begin{equation}
    \pi(s_t) = \argmax_{a\in\mathcal{A}(\mathbf{s_t})} \widehat{Q}(\mathbf{s_t}, a, \theta)
\label{eq:greedy-policy}
\end{equation}

Given an observed transition $(\mathbf{s_t}, a_t, r_{t+1}, \mathbf{s_{t+1}})$ from the dataset $\mathcal{D}$, the $Q$-model is trained to minimize the Bellman loss~\cite{kumar2020conservative}.
\begin{equation}
    \mathcal{L}_{Bellman} = \mathbb{E}_{\mathbf{s_t}, a_t, \mathbf{s_{t+1}}\sim\mathcal{D}} \left[\left(y - \widehat{Q}(\mathbf{s_t}, a_t, \theta)\right)^2\right]
\label{eq:bellman}
\end{equation}
where the $Q$-learning target $y$ is defined by:
\begin{equation}
    y = 
    \begin{cases}
    r_{t+1} &\;\; (\mathbf{s_{t+1}} \;\text{is a terminal state}) \\
    r_{t+1} + \gamma \max_a\widehat{Q}(\mathbf{s_{t+1}}, a) &\;\; (\text{otherwise})
    \end{cases}
\label{eq:q-learning-target}
\end{equation}

Since we consider our RL-based recommendation in an offline RL paradigm, we follow a state-of-the-art Conservative Q-Learning (CQL) framework to formulate our RL training target~\cite{kumar2020conservative}.
\begin{multline}
    \mathcal{L} = \alpha\mathbb{E}_{\mathbf{s_t}\sim\mathcal{D}}\left[\log\sum_a\exp(\widehat{Q}(\mathbf{s_t}, a))\right]-\alpha\mathbb{E}_{\mathbf{s_t}, a_t\sim\mathcal{D}}\left[\widehat{Q}(\mathbf{s_t}, a_t)\right] \\
    + \dfrac{1}{2}\mathcal{L}_{Bellman}
    \label{eq:new-loss}
\end{multline}
where $\alpha$ represents a weighting constant, and $\theta$ is omitted for clarity in the representation.

The rest of this section presents our proposed reward function and the design of replay memory, aimed at mitigating the issue of popularity bias. 
\subsubsection{Reward Function}
Consider a training project $\mathbf{u}$ with historical library usage $\mathcal{I}_u^*\subset\mathcal{I}$. A transition $(\mathbf{s}_t, a_t, r_{t+1}, \mathbf{s}_{t+1})$ can be defined from $\mathcal{I}_u^*$ as follows.
\begin{itemize}
    \item Current state: 
    $\mathbf{s}_t = Agg(\mathcal{I}^t_u)$ where $\mathcal{I}^t_u\subset\mathcal{I}_u^*$
    following Equation~\ref{eq:cold-start-repr}
    \item Action: $a_t\in\mathcal{I}_u^* \backslash\mathcal{I}^t_u$
    \item Reward: $r_{t} = 1 + \mathbf{e_{a_t} \cdot s_t}$ where $\cdot$ denotes the scalar product of two vectors. This design ensure positive reward values for observed interactions.
    \item Next state: $\mathbf{s}_{t+1} = Agg(\mathcal{I}_u^t\cup a_t)$
\end{itemize}

Using Equation~\ref{eq:weighted-repr}, the formula of the reward function is as follows.
\begin{align}
    \mathbf{r}_{t} &= \mathbf{e_{a_t} \cdot s_t} = \mathbf{e_{a_t}}\cdot Agg(\mathcal{I}^t_u) =\mathbf{e_{a_t}} \cdot \frac{1}{|\mathcal{I}^t_u|}\sum_{i\in\mathcal{I}^t_u}\mathbf{p_i} \\
    &= \frac{1}{|\mathcal{I}^t_u|} \sum_{i\in\mathcal{I}^t_u} 
    \mathbf{e_{a_t}} \cdot (\lambda\dfrac{\sum_{u\in\mathcal{U}^i}y(u, i)e_{u}}{\sum_{u\in\mathcal{U}^i}y(u, i)} + (1-\lambda)\mathbf{e_i}) \\
    &= \frac{1}{|\mathcal{I}^t_u|} (\lambda\sum_{i\in\mathcal{I}^t_u} 
    (\dfrac{\sum_{u\in\mathcal{U}^i}y(u, i) \mathbf{e_{a_t}} \cdot e_{u}}{\sum_{u\in\mathcal{U}^i}y(u, i)}) + 
    (1-\lambda)\sum_{i\in\mathcal{I}^t_u} \mathbf{e_{a_t}} \cdot \mathbf{e_i}))
    \label{eq:pb-reward}
\end{align}

In the final formula, the term $\mathbf{e_{a_t}} \cdot \mathbf{e_u}$ can be interpreted as a \textit{vote} from a known project $\mathbf{e_u}$ that shares common interests with current user induced by $\mathcal{I}^t_u$ via library $a_t$, which can be viewed as an implicit form of CF. Furthermore, the influence of popular libraries on the reward \(\mathbf{r}\) is adjusted through normalization using the denominator in Equation~\ref{eq:weighted-repr}.

\subsubsection{Popularity-Aware Replay Buffer}
The RL agent is trained over multiple epochs, with each epoch involving a full pass through all training projects. For each historical project-library usage associated with a project \( u \) in the training set, a unique set of transitions is generated in each epoch, representing a single recommendation step for \( u \). This approach enables the recommendation policy to be trained across diverse contexts. We propose a sampling process to generate transitions by sampling states and actions from the historical library usage $\mathcal{I}^*_u$. 

At each training step, a batch of transitions is drawn from the replay memory \( \mathcal{M} \), which retains past interactions to update the \( Q \)-network~\cite{van2016deep}. This enables the agent to learn from previous experiences, enhancing its generalization capability.
However, long-tailed libraries a low probability of being included in \( \mathcal{M} \) due to their rare selection in random action sampling. This limitation reduces the model's ability to recommend less popular libraries to relevant projects when appropriate.
To address popularity bias, we partition \( \mathcal{M} \) into three subsets: the rare partition \( \mathcal{M}_{rare} \), the random partition \( \mathcal{M}_{rand} \), and the sequential partition \( \mathcal{M}_{seq} \). We define the size ratios as \( \mu_{rare} = |\mathcal{M}_{rare}| / |\mathcal{M}| \), with similar definitions for \( \mu_{rand} \) and \( \mu_{seq} \), ensuring that $\mu_{rare}+\mu_{rand}+\mu_{seq} = 1$. 
As the names imply, the rare partition primarily stores long-tailed interactions, the random partition ensures gradient updates by incorporating both past and recent interactions through a random sampling strategy, and the sequential partition focuses on new projects and fresh interactions, employing a Round-Robin approach.
The sampling process guarantees that candidates from each partition are selected in proportion to their respective \( \mu \) values. 
We redefine the training loss function to account for the three partitions.
\begin{equation}
    \mathcal{L} = \mu_{rare}\mathcal{L}_{rare} + \mu_{rand}\mathcal{L}_{rand} + \mu_{seq}\mathcal{L}_{seq}
\label{eq:weighted-L}
\end{equation}
where in \eqref{eq:weighted-L}, $\mathcal{L}_{rare}$, $\mathcal{L}_{rand}$, $\mathcal{L}_{seq}$ denote the loss calculated by \eqref{eq:new-loss} over $\mathcal{M}_{rare}$, $\mathcal{M}_{rand}$ and $\mathcal{M}_{seq}$, respectively. 



\begin{figure}[t]
    \centering
    \includegraphics[width=0.5\textwidth]{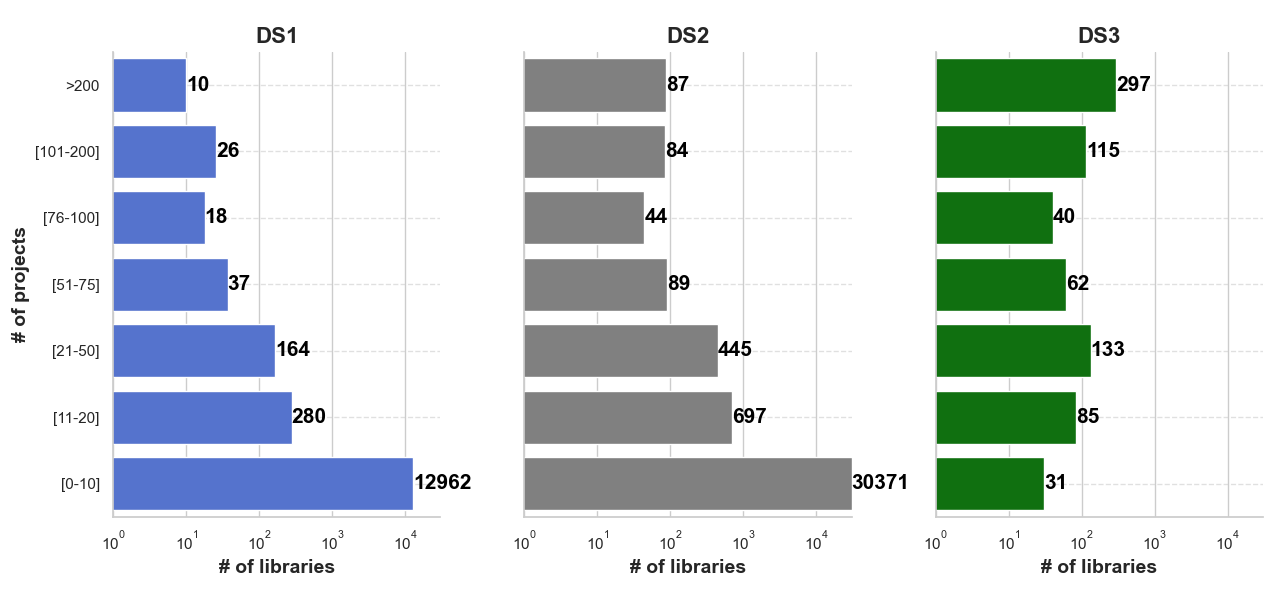}
    \caption{The distribution of libraries across three datasets.}
    \label{fig:dataset}
\end{figure}

\section{Experiment Setup}
\label{sec:settings}
In this section, we present the empirical evaluation and settings conducted to study the performance of our proposed approach.

\subsection{Benchmark Datasets}

In this study, we utilize a benchmark dataset from previous research~\cite{nguyen2020crossrec,he2020diversified,nguyen2023dealing}, comprising three Java open-source software projects. Among them, DS1 and DS2 correspond to generic software systems, while DS3 is specifically related to Android applications. Table~\ref{tab:dataset} provides a statistical summary of these datasets. Notably, all three datasets exhibit a long-tail distribution, particularly DS1, where 10,212 out of 13,497 libraries appear only once throughout the entire dataset.
Figure~\ref{fig:dataset} shows the distribution of libraries across three datasets. It can be observed that in DS1, only 10 libraries are highly popular by being included in more than 200 projects, whereas this number increases to 87 for DS2 and 297 for DS3. 
Additionally, a significant number of libraries--30,371 in DS2--are used in fewer than 10 projects. DS3 exhibits a different pattern, with a significantly high number of projects, i.e., 56,091 projects but only 762 libraries. We observe that although this dataset does not exhibit a strongly long-tail distribution, it contains a remarkably high number of interactions, with 537,011 project-library usage records.



\subsection{Evaluation Metrics}
\begin{table}[t]
    \centering
    \caption{TPL Dataset details.}
    \begin{tabular}{lccc}
        \hline
        \textbf{Dataset} & \textbf{DS$_1$} & \textbf{DS$_2$} & \textbf{DS$_3$} \\
        \hline
        \# projects & 1,200 & 5,200 & 56,091 \\
        \# libraries & 13,497 & 31,817 & 762 \\
        \hline
    \end{tabular}
    \label{tab:dataset}
\end{table}
We assess the performance of TPL recommendation by ranking the top-\( K \) libraries generated by the system and comparing them against the ground-truth data. This evaluation utilizes widely adopted metrics in recommender system, including \texttt{Precision@K} and \texttt{Recall@K}. Additionally, to measure the effectiveness in reducing popularity bias, we employ \texttt{EPC@K} (Expected Popularity Complement) and \texttt{Coverage@K}. The definitions of these metrics are provided in previous studies~\cite{zhang2022incorporating,nguyen2023dealing}. Due to space constraints, we omit these details.
By default, we set \( K = 10 \) and use 10-fold cross-validation to train and evaluate the system on the benchmark dataset. The final results are reported as the average values of the aforementioned metrics.



\subsection{Parameter Settings} 
We adopt the original configuration of the \LGCN model~\cite{he2020lightgcn}, utilizing two embedding layers for the encoder. The model is trained with a batch size of 1024, a learning rate of \(1 \times 10^{-4}\), and an \(L_2\) regularization coefficient of \(1 \times 10^{-5}\). The embedding dimension is set to 64. To optimize \LGCN, we employ \texttt{BCLoss} with 128 negative samples. Training stops if \(Recall@10\) does not improve after 20 consecutive epochs.


We implemented our RL-based recommender using a double and dueling DQN architecture~\cite{van2016deep,wang2016dueling}, setting the hidden layer dimension to 256.
Our RL agent is trained with a learning rate of \(1\times10^{-3}\) using a Cosine Annealing scheduler \cite{loshchilov2016sgdr} for 20 epochs. The weight parameters are set as \(\alpha = 5.5\) and \(\lambda = 0.5\).
For replay buffer partition, we set $\mu_{rare} = 0.2$, $\mu_{rand} = 0.5$ and $\mu_{seq} = 0.3$. 
We define the popularity rate of an action as the proportion of projects that have interacted with a given library relative to the total number of projects in the training dataset. Actions are classified as {\it rare} if their popularity in the training set is below $0.1$, while {\it popular} actions have a popularity rate exceeding $0.9$.
\subsection{Research Questions}
\label{sec:rq}
In this research we aim to answer to following research questions.

\vspace{.2cm}
\noindent \rqone~
This research question aims to assess the effectiveness of our approach in handling cold-start recommendations. We compare its performance with the recent work CrossRec~\cite{nguyen2020crossrec}. In the general context of recommendation systems, the cold-start scenario arises when either a new user or a new item enters the system. For TPL recommendation, previous research has primarily focused on handling new or recently created projects~\cite{li2021embedding,nguyen2023dealing}. Therefore, to evaluate our approach, we partition each dataset into two subsets: training projects \(\mathcal{U}_{train}\) and testing projects \(\mathcal{U}_{test}\), ensuring that \(\mathcal{U}_{train} \cap \mathcal{U}_{test} = \varnothing\).
Each set of projects is provided along with its complete list of project-library interactions. We train the recommendation with \(\mathcal{U}_{train}\) and employ \(\mathcal{U}_{test}\) for prediction.

For a project \( u \in \mathcal{U}_{test} \), its corresponding set of observed interactions \( \mathcal{I}_u \) is further divided into a query set \( \mathcal{I}^q_u \) and a test set \( \mathcal{I}^t_u \), ensuring that \( \mathcal{I}^q_u \cap \mathcal{I}^t_u = \varnothing \).
During the testing phase, the RL agent is provided with \( \mathcal{I}^q_u \) and leverages this prior knowledge to generate relevant recommendations, which are then evaluated by comparing them against \( \mathcal{I}^t_u \).
It is important to highlight that our approach assumes cold-start projects have at least some interactions. As a result, achieving an absolute cold-start recommendation scenario (i.e., \( \mathcal{I}^q_u = \varnothing \)), where no additional item-independent information about the new user \( u \) is available, is not feasible. This setting is motivated by the fact that, in general recommendation systems, metadata such as user preferences, behaviors, or item attributes (e.g., descriptions, value, price) are typically available and utilized to address the cold-start problem. However, in the context of TPL recommendation, such metadata is not accessible. Instead, a portion of the new project's interactions is often provided to offer some prior knowledge, enabling meaningful recommendations. 
Since library metadata is not available, any libraries that appear in the test set but are missing from the training set are excluded.
To further evaluate the performance of \tool, we impose a more challenging cold-start scenario by reducing the size of \( \mathcal{I}^q_u \) to just 30\% of the original splits defined in \cite{nguyen2023dealing}, while transferring the remaining 70\% to \( \mathcal{I}^t_u \). We apply this setting to both \tool and \CR, ensuring that the two benchmark datasets, DS1 and DS2, are split in the same manner.

\vspace{.2cm}
\noindent \rqtwo~
In this research question, we evaluate the effectiveness of our proposed approach in mitigating popularity bias by comparing its performance with the baseline method \CR~\cite{nguyen2020crossrec}. Inspired by the prior work of Nguyen et al.~\cite{nguyen2023dealing}, we adopt the same settings as in the cold-start recommendation scenario, where the entire set \( \mathcal{I}^q_u \) is used as prior knowledge about the testing projects. 

\vspace{.2cm}
\noindent \rqthree~
In this research question, we examine the effectiveness of \tool in a traditional interaction-split setting, where the entire interaction matrix is divided into training and testing sets. This differs from the previous setting, where only the set of users (projects) was partitioned. Since \LGCN~\cite{he2020lightgcn} is specifically designed for interaction-split recommendation, we explore its effectiveness when integrated into our RL framework. To achieve this, we train the \LGCN encoder under two settings: with and without \texttt{BCLoss}, resulting in two variants of \tool—\RLGCN and \RLBC. We then compare these variants with 
\GR~\cite{li2021embedding}, which is also designed for interaction-split TPL recommendation.



For each project \( u \in \mathcal{U} \), its interaction history \( \mathcal{I}_u \) is split into two disjoint subsets: \( \mathcal{I}_u^{train} \) and \( \mathcal{I}_u^{test} \). 
During training, the model is exposed to the interaction behaviors of all projects \( u \in \mathcal{U} \) but has access only to \( \mathcal{I}_u^{train} \). 
The test set \( \mathcal{I}_u^{test} \) serves as ground truth for evaluating the generated recommendations. To evaluate model performance, we adopt the interaction-split configuration from Nguyen et al.~\cite{nguyen2023dealing} for partitioning DS1, DS2, and DS3. The reported results of \GR on DS2 and DS3 from this study are used for comparison with our proposed approach.


\section{Results and Discussion}
\label{sec:rs}

\subsection{\rqone}

Table~\ref{tab:result_coldstart} and~\ref{tab:result_coldstart_ex}  report the performance of \tool in terms of \texttt{Precision@10} and \texttt{Recall@10}, comparing it to \CR under the cold-start splitting method introduced in \cite{nguyen2023dealing} for DS1 and DS2. 
We analyze two cold-start recommendation scenarios: one where the entire interaction subset \( \mathcal{I}^q_u \) is utilized (Table~\ref{tab:result_coldstart}) and another where only 30\% of \( \mathcal{I}^q_u \) is considered (Table~\ref{tab:result_coldstart_ex}) (See Section~\ref{sec:rq}).
As shown in Table~\ref{tab:result_coldstart}, \tool demonstrates strong performance in the cold-start user setting across both datasets, surpassing \CR by 42.46\% and 13.94\% in \texttt{Recall@10} on DS1 and DS2, respectively.
In terms of \texttt{Precision@10}, \tool achieves a 25.47\% improvement on DS1 and performs comparably to \CR on DS2.


\begin{table}[t]
    \centering
   \footnotesize	
    \caption{Effectiveness of \tool compared to \CR in addressing cold-start recommendation (using 100\% $\mathcal{I}^q_u$).}
    \begin{tabular}{l|cc|cc}
        \hline
        \multirow{2}{*}{\textbf{Model}} & \multicolumn{2}{c|}{\textbf{DS1}} & \multicolumn{2}{c}{\textbf{DS2}}\\
        \cline{2-5} & Precision@10& Recall@10& Precision@10& Recall@10\\
        \hline
        \textbf{\CR}~\cite{nguyen2020crossrec} & 12.33 & 28.97& 32.89 & 34.29 \\
        \textbf{\tool} & \textbf{15.47} & \textbf{41.27} &	\textbf{32.48}&	\textbf{39.07}\\
        \hline
    \end{tabular}      
\label{tab:result_coldstart}
\end{table}

\begin{table}[t]
    \centering
   \footnotesize	
    \caption{Effectiveness of \tool compared to \CR in addressing extreme cold-start recommendation (using 30\% $\mathcal{I}^q_u$).}
    \begin{tabular}{l|cc|cc}
        \hline
        \multirow{2}{*}{\textbf{Model}} & \multicolumn{2}{c|}{\textbf{DS1}} & \multicolumn{2}{c}{\textbf{DS2}}\\
        \cline{2-5} & Precision@10& Recall@10& Precision@10& Recall@10\\
        \hline
        \textbf{\CR}~\cite{nguyen2020crossrec} & 25.49&	{16.27}&	27.35&	{22.37}\\
        \textbf{\tool} & \textbf{30.9}&	\textbf{19.34}&	\textbf{38.56}&	\textbf{25.53}\\
        \hline
    \end{tabular}      
    \label{tab:result_coldstart_ex}
\end{table}

In order to demonstrate the effectiveness of the proposed cold-start representation, we examine a more challenging scenario where very little prior knowledge (only 30\% of $\mathcal{I}^q_u$) is considered for each new project. 
As shown in Table~\ref{tab:result_coldstart_ex}, \tool consistently outperforms \CR across all performance metrics in both datasets.
In terms of \texttt{Precision@10}, \tool demonstrates a notable improvement of 21.22\% on DS1 and 40.99\% on DS2. Similarly, for \texttt{Recall@10}, it achieves an increase of 18.87\% on DS1 and 14.13\% on DS2, maintaining consistent performance across both datasets.



\subsection{\rqtwo}
We assess the effectiveness of \tool in mitigating popularity bias using two benchmark datasets, DS1 and DS2. Following the experimental setup of Nguyen et al.~\cite{nguyen2020crossrec}, we apply 10-fold cross-validation for model evaluation and report the average performance on two key metrics: \texttt{EPC@10} and \texttt{Coverage@10}.
The result is reported in Table \ref{tab:result_popularity}. Overall, \tool demonstrates strong performance across both metrics, outperforms \CR by 17.64\% and 5.58\% in \texttt{EPC@10} on DS1 and DS2, respectively.
In terms of \texttt{Coverage@10}, both methods exhibit comparable performance.
Recent research on mitigating popularity bias in TPL recommendation by Nguyen et al.~\cite{nguyen2023dealing} has shown that while certain methods (e.g., \CR, which is also examined in this study) enhance the diversity of recommendation results, particularly for rare items, they often come at the cost of reduced prediction accuracy.
The results of \texttt{Precision@10} and \texttt{Recall@10} can be observed from Table~\ref{tab:result_coldstart} as we have used the same configuration for experiment.
It is evident that \tool not only effectively mitigates the impact of popularity bias but also preserves prediction accuracy.

\begin{table}[t]
    \centering
   \footnotesize	
    \caption{Effectiveness of \tool compared to state-of-the art baseline models in addressing popularity bias.}
    \begin{tabular}{l|cc|cc}
        \hline
        \multirow{2}{*}{\textbf{Model}} & \multicolumn{2}{c|}{\textbf{DS1}} & \multicolumn{2}{c}{\textbf{DS2}}\\
        \cline{2-5} & EPC@10& Cov.@10 &  EPC@10& Cov.@10\\
        \hline
        \textbf{\CR}~\cite{nguyen2020crossrec} &  43.43 & \textbf{2.99}	 &  58.08 & 1.67\\
        \textbf{\tool}  &
        \textbf{51.09}& 2.4 &  \textbf{61.32}& \textbf{2.01}\\
        \hline
    \end{tabular}      
    \label{tab:result_popularity}
\end{table}

\subsection{\rqthree}
Table \ref{tab:interaction-split} presents the performance of \tool under a traditional interaction-split configuration for recommendation. We adopt the same experimental setting for our proposed approach while reusing the reported results of \GR~\cite{li2021embedding}. To assess the contribution of \LGCN embeddings, we evaluate two variants of \tool: one incorporating BCLoss and the other without it.

We observe that \RLBC, the default configuration of \tool, demonstrates a strong performance across all datasets in terms of prediction accuracy metrics including \texttt{Precision@10} and \texttt{Recall@10}.
In DS2, which exhibits a highly long-tail distribution of rare libraries, \RLBC significantly outperforms \RLGCN, achieving improvements of 806\% in \texttt{Precision@10} and 904\% in \texttt{Recall@10}. This demonstrates that \texttt{BCLoss} plays a crucial role in mitigating popularity bias, allowing to better capture the representation embeddings for both projects and libraries. 
Compared to \GR, \RLBC achieves superior performance on \texttt{Precision@10} and \texttt{Recall@10}, while demonstrating comparable results on \texttt{EPC@10} and \texttt{Coverage@10}.
We conclude that, despite being designed for cold-start recommendation, the proposed RL agent is still capable of delivering strong performance in the traditional interaction-split setting.

\begin{table}[t]
    \centering
    \footnotesize
    \caption{Experimental results on an interaction-split recommendation setting}
    \label{tab:interaction-split}
    \renewcommand{\arraystretch}{1.2}
    \begin{tabular}{l|l|cccc}
        \hline
        \textbf{Dataset} & \textbf{Model} & Prec.@10 & Rec.@10 & EPC@10 & Cov.@10 \\
        \hline
        \multirow{2}{*}{\textbf{DS1}}  
        & \textbf{\GR}~\cite{li2021embedding}     & - & - & - & - \\
        & \textbf{\RLGCN}  & 32.01 & 90.98 & 26.89 & 20.83 \\
        & \textbf{\RLBC}   & \textbf{32.35} & \textbf{92.57} & \textbf{28.35} & \textbf{37.15} \\
        \hline
        \multirow{3}{*}{\textbf{DS2}} 
        & \textbf{\GR}~\cite{li2021embedding}     & 6.32 & 21.06 & \textbf{70.56} & 7.73 \\
        & \textbf{\RLGCN}  & 1.21 & 4.06 & 48.30 & 7.69 \\
        & \textbf{\RLBC}   & \textbf{10.97} & \textbf{40.79} & 69.91 & \textbf{8.64} \\
        \hline
        \multirow{3}{*}{\textbf{DS3}} 
        & \textbf{\GR}~\cite{li2021embedding}    & 20.98 & 69.94 & 46.22 & \textbf{58.45} \\
        & \textbf{\RLGCN}  & 20.12 & 67.07 & 42.59 & 40.10 \\
        & \textbf{\RLBC}   & \textbf{22.07} & \textbf{73.57} & \textbf{48.43} & 56.82 \\
        \hline
    \end{tabular}
\end{table}

\section{Related Work}
\label{sec:related-works}
\textbf{Popularity bias in TPL recommendation.} According to Nguyen et al.~\cite{nguyen2023dealing}, the issue of popularity bias in TPL recommender systems has remained largely unexplored within the SE community. Of the studies reviewed, LibSeek ~\cite{he2020diversified} stands as the only 
attempt to improve the diversity of recommendation results by introducing an adaptive weighting mechanism. Other TPL recommender systems are prone to popularity bias due to various factors. LibRec ~\cite{thung2013automated} tends to retrieve popular items due to uses of association mining~\cite{agrawal1993mining}. CrossRec~\cite{nguyen2020crossrec}
relies on similar users to guide the recommender. GRec~\cite{li2021embedding} uses a graph-based link prediction formulation, which can also lead to popularity bias due to lack of 
information over rare items.

\textbf{RL for recommender systems.} According to ~\cite{afsar2022reinforcement}, RL-incorporated recommender system can be divided into RL-based~\cite{chang2021music, wang2020hybrid, kokkodis2021demand} and Deep RL (DRL)-based methods~\cite{ge2021towards, zhang2019text, zhao2018recommendations}. DRL-based methods can be further classified into value-based~\cite{zhao2018recommendations}, policy gradient~\cite{zhang2019text} and actor-critic methods~\cite{ge2021towards}. \tool can be classified into value-based DRL recommender system.

\section{Threats to Validity}
\label{sec:threat}
The primary threats to validity in our study relate to performance comparisons between \tool and benchmark frameworks. To mitigate potential biases from manual implementations, we utilize the data splits and corresponding benchmark results for \CR and \GR provided by \cite{nguyen2023dealing}. To maintain consistency in result computation, we re-implemented the evaluation metrics and applied this evaluation kit uniformly across all tasks. Another key consideration is the generalizability of \tool across different recommendation tasks and scales. We address this partially by assessing its robustness under varying proportions of known interactions in cold-start user settings, with results presented in Table~\ref{tab:result_coldstart_ex}.

\section{Conclusion}
\label{sec:conclusion}
In this paper, we introduced \tool, an RL-based framework for TPL recommendation designed to address both popularity bias and cold-start challenges. Extensive experiments on widely used TPL datasets demonstrated that \tool achieves competitive performance while effectively mitigating popularity bias. These results highlight the potential of integrating RL into recommender systems and enhancing CF models for cold-start scenarios. In future work, we aim to further refine the components of \tool to enhance its overall performance and effectiveness.

\begin{acks} 
	
	This paper has been partially supported by the MOSAICO project (Management, Orchestration and Supervision of AI-agent COmmunities for reliable AI in software engineering) that has received funding from the European Union under the Horizon Research and Innovation Action (Grant Agreement No. 101189664). The work has been partially supported by the EMELIOT national research project, which has been funded by the MUR under the PRIN 2020 program (Contract 2020W3A5FY). It has been also partially supported by the European Union--NextGenerationEU through the Italian Ministry of University and Research, Projects PRIN 2022 PNRR \emph{``FRINGE: context-aware FaiRness engineerING in complex software systEms''} grant n. P2022553SL. We acknowledge the Italian ``PRIN 2022'' project TRex-SE: \emph{``Trustworthy Recommenders for Software Engineers,''} grant n. 2022LKJWHC. 
\end{acks}

\bibliographystyle{ACM-Reference-Format}
\bibliography{sample-base}


\end{document}